\documentclass[aps,prl,twocolumn,groupedaddress,showpacs]{revtex4}
\usepackage{graphicx}
\usepackage{epsfig}
\begin{document}

%\preprint{}

\title{Ion-ion correlation and charge reversal at titrating solid interfaces}
%\title{Ion-ion correlation and charge reversal of titrating silica particles in divalent electrolytes}


\author{Christophe Labbez$^1$, Bo J\"onsson$^2$, Michal Skarba$^3$ and Michal Borkovec$^3$}
% \email[]{Christophe.Labbez@u-bourgogne.fr}
%\thanks{}
\affiliation{$^1$ Institut Carnot de Bourgogne, UMR 5209 CNRS, Universit\'e de Bourgogne, F-21078 Dijon, France\\
$^2$ Department of Theoretical Chemistry, Chemical Center, Lund University, POB 124, S-221 00 Lund, Sweden\\
$^3$ Department of Inorgarnic, Analytical and Applied Chemistry, Geneva University, 1211 Geneva 4, Switzerland}
% \author{Bo J\"onsson}
% \email[]{Bo.Jonsson@teokem.lu.se}
%\thanks{}
% \affiliation{Department of Theoretical Chemistry, Chemical Center, Lund University, POB 124, S-221 00 Lund, Sweden}

% \author{Michal Skarba and Michal Borkovec}
% \affiliation{Department of Inorgarnic, Analytical and Applied Chemistry, Geneva University, 1211 Geneva 4, Switzerland}

\date{\today}

\begin{abstract}
Confronting grand canonical titration Monte Carlo simulations (MC) with recently published titration and charge reversal (CR) experiments on silica surfaces by Dove {\it et al.} \cite{Dove:05} and van der Heyden {\it et al.} \cite{Heyden:06}, we show that ion-ion correlations quantitatively explain why divalent counterions strongly promote surface charge which, in turn, eventually causes a charge reversal (CR). Titration and CR results from simulations and experiments are in excellent agreement without any fitting parameters. This is the first unambiguous evidence that ion-ion correlations are instrumental in the creation of highly charged surfaces and responsible for their CR. Finally, we show that charge correlations result in ``anomalous'' charge regulation in strongly coupled conditions in qualitative desagreement with its classical treatment.
\end{abstract}

%\keywords{}

\maketitle

\section{}
A key parameter for understanding electrostatic interactions in charged suspensions is the surface charge density, $\sigma$ \cite{Honig:95,Hunter:79}. 
This parameter may control phase separation, phase transition \cite{Chang:95}, promote dissolution \cite{Dove:005}, control particle growth \cite{Dove:06}, 
colloidal stability \cite{Derjaguin:41,Verwey:48} and membrane selectivity \cite{Leung:06} and play a significant role in catalysis and 
fluid transport through ionic channels \cite{Stein:04, Boda:07}. Typically a surface becomes charged through a titration process where
surface groups ionize, {\it e.g.} the titration of silanol groups -Si-OH $\rightarrow$ -Si-O$^-$ + H$^+$.

Ion-ion correlation in solutions of charged colloids has attracted much interest due to its ability to generate net attractive 
interaction between equally charged colloids. This is, however, only one aspect of ion-ion correlation and an equally important aspect is
its facilitation of surface titration processes. An example of vital importance is the surface titration of calcium silicate hydrates
in cement paste, which reach a very high surface charge density due to ion-ion correlation among the calcium counterions. The high
surface charge density is then responsible for the setting of cement \cite{Delville:97,Jonsson:05,Labbez:06}.

An additional effect, due to ion-ion correlation, is when a charged surface immersed in a multivalent electrolyte more
than compensates its charge resulting in a \textit{charge reversal} (CR) \cite{Pashley:84b,Sjostrom:96b}. 
That is, the surface attracts counterions in excess 
of its own nominal charge and the apparent charge seen a few {\AA}ngstr\"om from the surface appears to be of opposite sign to the 
bare surface. CR is experimentally characterized by several methods, {\it e.g.} streaming current \cite{Heyden:06} and can induce 
repulsion between oppositely charged surfaces \cite{Besteman:04,Trulsson:06}. Despite a vast number of studies the origin of charge 
reversal is still controversial as highlighted in a recent review article \cite{Lyklema:06}, where Lyklema points out the lack of 
discrimination between the \textit{chemical} (specific chemical adsorption) and \textit{physical} (Coulombic interactions 
including charge correlations) origin of CR.

The charging of interfaces (mineral or organic) has traditionally been described by the {\it surface complexation model} (SC) \cite{Hiemstra:89, Sverjensky:93, Borkovec:97},
where titration of surface groups is described by an ensemble of mass balance equilibria with associated equilibrium constants. 
The SC is based on the Poisson-Boltzmann (PB) equation augmented with a so-called Stern layer. The PB equation, however, neglects ion-ion correlation and {\it e.g.} fails to predict the
attraction between equally charged surfaces \cite{Guldbrand:84, Zohar:06} as well as the repulsion between oppositely charged 
ones \cite{Trulsson:06}. Despite its widespread use the limitations of the SC model in highly coupled systems have never been examined. 
Moreover, although multivalent ions are ubiquitous in many systems, there are only few experimental studies
with titration of metal oxide particles in the presence of such ions \cite{Dove:01, Dove:05, Foissy:94}.

Recently a grand canonical titration simulation method was developed \cite{Labbez:07} and applied to a calcium silicate hydrate (C-S-H) 
solution \cite{Labbez:06}. An excellent agreement was found between experiment and simulation for the titration process as well as for the CR. 
Unfortunately, due to the solubility range of C-S-H, the full titration curve is not accessible experimentally.

In the present letter, we scrutinize the prediction from our MC simulations for the charging process and CR by
comparing to experimental data for silica particles dispersed in sodium and calcium chloride solutions \cite{Dove:05}.
In particular, we focus on the contribution of the correlations (ion-ion and ion-site).
We also present a simulated charge regulation of two parallel silica surfaces in a Ca$^{2+}$ salt solution when decreasing 
their separation. The experimental investigation of surface titration of silica fume particles
by Dove {\it et al.} \cite{Dove:05} and CR by multivalent counterions from a streaming current analysis in silica 
nanochannels by van der Heyden {\it et al.} \cite{Heyden:06} will serve as references. 

The simulations are based on the Primitive Model with the solvent as a structureless medium characterized by its relative 
dielectric permittivity, $\epsilon_r$.  All charged species are treated as charged hard spheres and the interaction, between 
two charges \textit{i} and \textit{j} separated a distance \textit{r} is, 
\begin{equation}
u(r) = \frac{Z_i Z_j e^2}{4 \pi \epsilon_0 \epsilon_r r} \hspace{0.4cm} \mbox{if}  \hspace{0.4cm} r>d_{hc} \\
\label{Coulomb}
\end{equation}
otherwise $u(r) = \infty$.
$Z_i$ is the ion valency, $e$ the elementary charge, $\epsilon_0$ the dielectric permittivity of vacuum and $d_{hc}$  
the hard sphere diameter of an ion always equal to 4 {\AA}. The particles are modeled as infinite planar walls with
explicit titratable sites with the intrinsic dissociation constant $K_0$ distributed on a square lattice. The titration process can be described as,
\begin{equation}
MOH \rightleftharpoons MO^-+H^+, \hspace{0.4cm} K_0 = \frac{a_M a_H}{a_{MOH}}
\label{Titration}
\end{equation}
where the $a$'s denote the activities of the species.
The model %, schematically represented in Fig.\ref{Model},
 is solved using a Grand Canonical Monte Carlo 
method  \cite{Frenkel}, {\it i.e.} at constant pH and chemical potential of the ions. In addition to 
ordinary MC moves the titratable sites are allowed to change their charge status. We imagine the deprotonation of a surface 
site as a two-step process: the release of a proton from the surface followed by an exchange of an ion pair ($H^+$, $B^-$) 
with the bulk. The corresponding Boltzmann factor for the trial energy can be expressed as,
\begin{equation}
\exp(-\beta \Delta U) = \frac{N_B}{V} \exp[-\beta (\mu_B + \Delta U^{el}) +\ln 10 (\mathrm{pH}-\mathrm{p}K_0)]
\label{DeProt}
\end{equation}
where $\mu$ represents the chemical potential of a particular ion, $V$ is the volume, $N_B$ the number of ions $B^-$ and 
$\Delta U^{el}$ the change in electrostatic energy. An analogous expression holds for protonation, 
for more details see \cite{Labbez:06}. 

\begin{figure}[h,t]
\begin{center} \includegraphics[width=8cm]{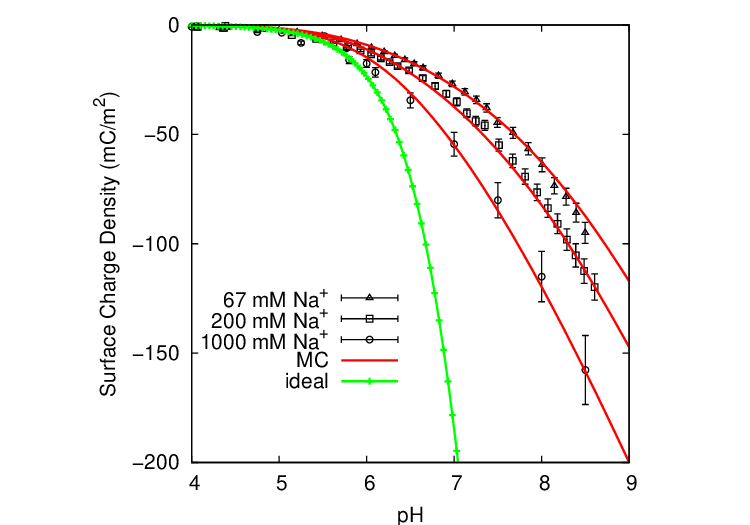}\end{center}
\caption{Simulated (MC) and experimental (symbols) \cite{Dove:05} surface charge density versus pH for silica particles dispersed 
in a sodium (1:1) salt solution at different concentrations. The ideal titration curve (ideal) is given as a reference}
\label{DoveComp}
\end{figure}

\noindent To establish the model parameters for a silica/solution interface, we turn our attention to the case of silica in 1:1
salt solution. The following set of parameters: $\mathrm{p}K_0$ = 7.7, $\rho_s$ = 4.8/nm$^2$, the surface site density, 
and $d_{is} =$ 3.5 {\AA}, the minimum
separation between an ion and a surface site, have been found to give a perfect description of the charging process,
see Fig.\ref{DoveComp}. The surface charge density, $\sigma$ increases, in absolute value, with pH, but is always smaller than 
in the ideal case due to the strong electrostatic repulsion between the deprotonated silanol groups. A change in the salt 
concentration modulates the screening of the electrostatic interactions which, in turn, results in a change in $\sigma$.

\begin{figure}[h,t]
\begin{center} \includegraphics[width=8cm]{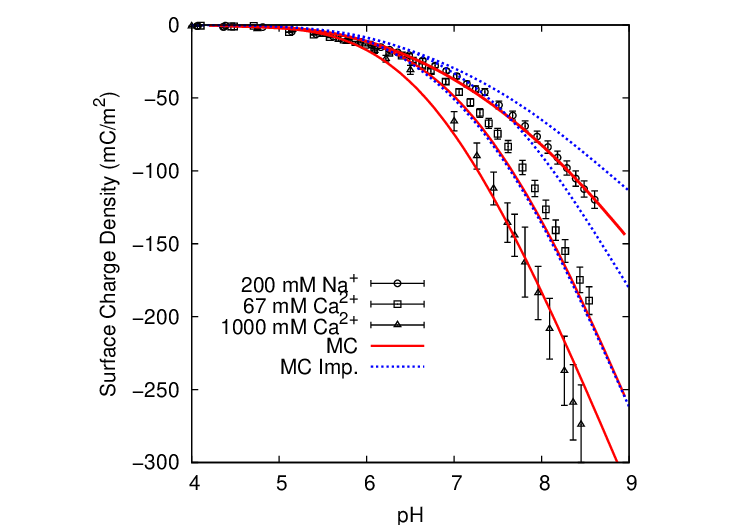}\end{center}
\caption{Simulated (MC) and experimental (symbols) \cite{Dove:05} surface charge density versus pH for silica particles dispersed
in a calcium (2:1) and sodium (1:1) salt solution at various
concentrations. The corresponding simulated titration
curves for particles having implicit sites (MC imp.) are also given
for comparison}
\label{DoveComp2}
\end{figure}

\noindent When the sodium counterions are replaced by divalent calcium ions, the simulations 
predict a strong increase in the surface charge density of silica in perfect agreement with the titration data of 
Dove {\it et al.} \cite{Dove:05} - see Fig.\ref{DoveComp2}. A similar behavior has been observed from titration experiments on 
titania particles \cite{Foissy:94} in multivalent electrolytes and, indirectly, from proton release measurements following adsorption of 
multivalent ions (or polyelectrolytes) \cite{Pittler:06} emphasizing the universality of the phenomenon. 
The classical SC approach fails to predict the effect, unless a complexation reaction of calcium with 
the silanol surface groups is invoked. Here, we unambiguously show that the charge correlations (ion-ion and ion-site),
without any additional fitting parameters, are enough to explain the increased silanol group ionization in the presence of
Ca$^{2+}$ ions. Thus, the complexation constants often invoked in the SC has no physical basis.

%{\bf In addition to the valence effect, Dove et al, also observed a ion specific effect. For silica in bivalent ions a ``reverse'' 
%lyotropic trend was found, i.e. the surface charge of silica is promoted in the order of increasing solvated ion radius 
%(decreasing crystallographic radius). Surprisingly, if one assume that the ions are dehydrated in contact with the silica surface 
%(then $d_Na \approx d_Ca \approx$ 1 {\AA} and $d_O \approx$ 1.6 {\AA}), the simulation prediction for $d_{si}$ ($\approx$ 1.5{\AA}) 
%is found to compare well with the theoretical value (1.3 {\AA}) and good predictions for the charging process of silica in the 
% %other bivalent cations is found, e.g. Mg$^{2+}$ in Fig.2. On the other hand, we should admit that $d_{si}$ is an 
%\textit{effective} distance that includes many other factors neglected in the Primitive Model.}

In an attempt to distinguish the effect of ion-ion from ion-site correlations a calculation was performed using a smeared out 
surface site density - see Fig.\ref{DoveComp2}.  A significant drop in $|\sigma|$ is observed, which becomes more pronounced 
when the counterion valence is high, although the qualitative behavior of $\sigma$ upon increasing the counterion valency remains
the same. Thus, the discrete nature of the surface groups have a quantitative but not qualitative effect and we can 
conclude that ion-ion correlations are more important than ion-site correlations, in contradiction with recent approximate
studies \cite{Pittler:06, Faraudo:07}. 

\begin{figure}[h,t]
\begin{center} \includegraphics[width=8cm]{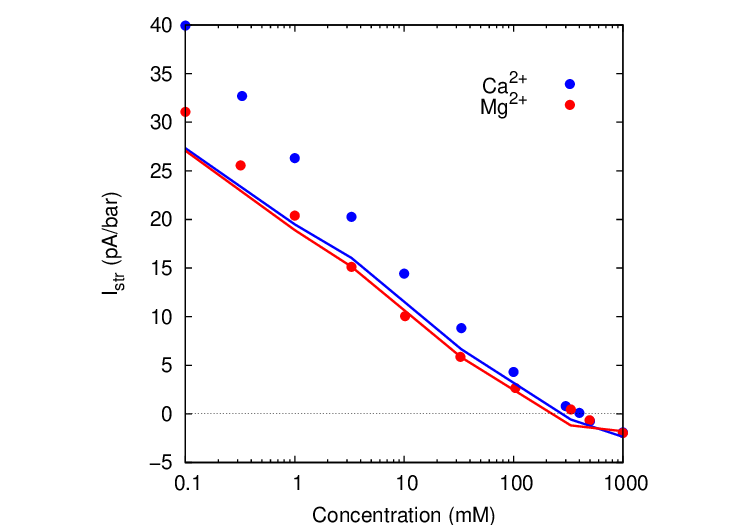}\end{center}
\caption{Comparison between simulated (line) and experimental (points) \cite{Heyden:06} streaming currents of a silica 
channel as a function of electrolyte concentration of a CaCl$_2$ and
MgCl$_2$ salt at pH 7.5. In the simulations, the site density is 4.8
nm$^{-2}$, p$K_0=7.7$, $d_{is}(\mathrm{Ca^{2+}})$ = 3.5 {\AA}, $d_{is}(\mathrm{Mg^{2+}})$ =
3.1 {\AA}.}
\label{Streaming}
\end{figure}

\noindent We can also compare our simulation data to the streaming current, $I_{str}$, experiments through silica channels
performed by van der Heyden {\it et al.} \cite{Heyden:06} in presence of di- and tri-valent counterions at pH 7.5. $I_{str}$ is computed from the 
simulated local charge density, $\rho(x)$. For a channel of height $h$ and width $w$ with $h>>w$, $I_{str}$ can be defined as,
\begin{equation}
I_{str}=2w\int_{0}^{h/2}\rho(X)v(X)dX \hspace{0.3cm} \mbox{with} \hspace{0.3cm} X=x-\frac{3}{2d_{hc}}
\label{StreamPlot}
\end{equation}
where $x$ is the distance normal to the silica surface and $v(x)$ is the local fluid velocity, approximated as a Poiseuille 
flow with boundary conditions: $v(X=0)=0$ and $v'(X=h/2)=0$. The zero velocity is arbitrarily set at $x=3/2d_{hc}$ where by 
definition we also find the electrokinetic potential. The simulations predict for both Mg$^{2+}$ and Ca$^{2+}$ a monotonic 
decrease in $I_{str}$ upon increasing salt concentration as also found experimentally - see Fig.\ref{StreamPlot}.
A sign reversal in the streaming potential is predicted at $c_{CR}\approx$ 310 mM for Ca$^{2+}$ and $c_{CR}\approx$ 280 mM 
for Mg$^{2+}$ in good agreement numbers observed experimentally \cite{Heyden:06}. What is more, the magnesium value is predicted to be
below that of calcium in agreement with experiments \cite{Heyden:06} due to a higher silica charge. Simulations were also performed 
for trivalent cations, where CR is predicted at sub-millimolar concentrations, $c_{CR} \approx 400 \mu$M, 
although times higher than what was observed with cobalt(III)sepulchrate \cite{Heyden:06}.

\begin{figure}[h,t]
\begin{center} \includegraphics[width=8cm]{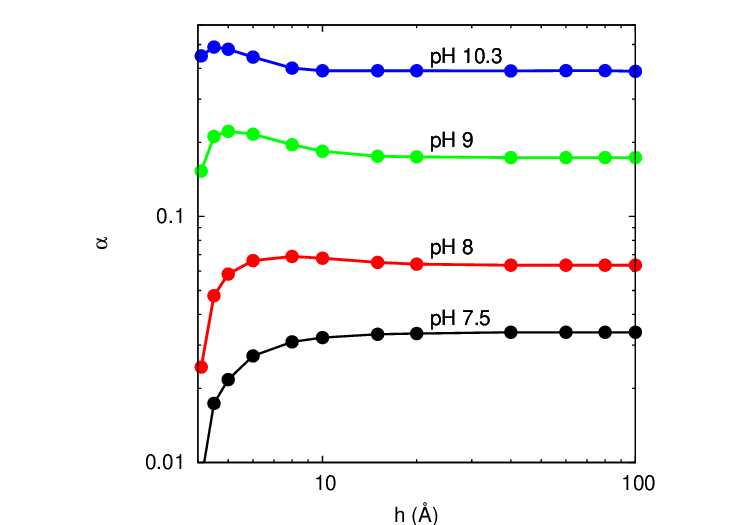}\end{center}
\caption{The degree of ionization, $\alpha$, as a function of surface separation at various pH values. The site density 
is 4.8 nm$^{-2}$ and p$K_0=7.7$ and the surfaces are in contact with a 0.2 mM CaX$_2$ solution.}
\label{ChargeRegulation}
\end{figure}

\noindent We now investigate the surface charging process of two parallel planar silica surfaces upon decreasing their 
separation, that is {\it the charge regulation}. The dependence of $\alpha$ on separation, $h$, in the presence of 0.2 mM Ca$^{2+}$ 
at various pH is shown in Fig. \ref{ChargeRegulation}. At low pH, a monotonic decrease of $\alpha$ with 
decreasing $h$, in conjunction with an increased osmotic repulsion (not shown), is found in qualitative agreement 
with previous mean-field studies \cite{Kirkwood:52,Ninham:71}. This result can be rationalized as a balance between 
entropy and energy. That is, when pH is well below p$K_0$, $\alpha$ is small and the free energy is dominated by the entropic 
term causing the particles to repel. As a result, the system minimizes its free energy by a reduction of the surface ionization. 
Conversely, atypical $\alpha$ vs $h$ curves are predicted for surfaces immersed in divalent electrolytes at large pH values 
($\mathrm{pH} \geq \mathrm{p}K_0$), where $\alpha$ is found to increase when shortening $h$. The degree of dissociation reaches a maximum at 
$h_{max}\approx$ 6 {\AA}, whereafter it decreases and annihilates at contact. The maximum in $\alpha$ corresponds exactly to 
the minimum of the potential of mean force (not shown). In this case, the energy (induced by the ion-ion correlations) dominates, 
which leads to an attraction between the particles and, subsequently, to an increase of the surface ionization for short $h$,
which in turn strengthen the attraction. Close to contact the entropy regains in importance and eventually 
causes $\sigma$ to collapse. However, when the Ca$^{2+}$ concentration is increased to more than 10 mM (not shown), the rise 
in $\alpha$ is dramatically strengthen while the drop at close to contact is hardly measurable. In other words, the surfaces retain 
and even increase their charges all the way to contact.

This charge regulation behaviour may partly explain why Ca$^{2+}$ promotes the experimentally observed dissolution of amorphous 
and crystalline silica surface under stress \cite{Israelachvili:06}. Indeed, as recently rationalized by Dove {\it et al.} 
\cite{Dove:005}, the dissolution proceeds through a nucleation process following the same mechanistic theory as developed for growth 
and it is shown to be promoted by surface ionization. 

Ion-ion correlation also seems to be relevant for membrane selectivity. As an example, we have calculated the L-type Ca-channel 
selectivity at various pH in a reservoir containing always 100 mM Na$^{+}$ to which is added either 1 mM or 100 $\mu$M Ca$^{2+}$. 
The Ca-channel was modeled as an infinite slit pore with $h$ = 9 {\AA}, with walls decorated with titratable carboxylic 
groups, $\mathrm{p}K_0$ = 4.8 and $\rho_{s}$ = 3/nm$^{2}$. The selectivity, $\xi$, was calculated as the ratio between the average calcium
and sodium concentrations. 
At pH 7 and 100 $\mu$M of Ca$^{2+}$ we found $\xi$=0.96 and $\alpha$ = 0.70, while at 1 mM Ca$^{2+}$ the values increased to 
$\xi$=3.7 and $\alpha$ = 0.84. 
Boda {\it et al.} \cite{Boda:07} tried to mimick the selectivity with a constant ionization of the -COOH groups, but were then forced
to decrease the relative dielectric permittivity of the protein to 10 in order to fit the experimental data. 
When pH is reduced to 5, $\xi$ drops to 0.10 and 0.15 at low and high calcium content, respectively
in agreement with the observed loss in membrane selectivity at low pH \cite{Delisle:00}.

To summarize, we have shown by confronting Monte Carlo simulations with independent experiments that the dominating interaction which controls the charging process and charge reversion of silica in multivalent  
electrolyte is of purely electrostatic origin and strongly dependent on ion-ion correlations. 
In particular, we have demonstrated that the charging process and charge reversal are intimately related. 
We have also demonstrated that the accepted view on charge regulation as a monotonic drop in $|\sigma|$ with decreasing 
separation is qualitatively wrong in highly coupled systems. Instead, $|\sigma|$ is found to increase with decreasing $h$.
This has profound influence for the stability of colloidal particles and for the early setting of normal Portland cement.

We thank R. Kjellander and S. G. Lemay for helpful discussions. M. Dove and F.H.J. van der Heyden are greatly acknowledge for 
the kind permission to use their experimental data.

\end{document}